\documentclass[journal]{IEEEtran}
\IEEEoverridecommandlockouts
\usepackage{mathrsfs}
\usepackage{amsfonts}
\usepackage{ifpdf}
\usepackage{cite}

\ifCLASSINFOpdf
\usepackage[pdftex]{graphicx}
\else \fi
\usepackage[cmex10]{amsmath}
\usepackage{array}
\usepackage{mdwmath}
\usepackage{mdwtab}
\usepackage{eqparbox}
\usepackage[tight,footnotesize]{subfigure}
\usepackage{algorithmic}
\usepackage{algorithm}
\usepackage{amsmath}
\usepackage{epsfig}
\usepackage{ae}
\usepackage{amssymb}
\usepackage{times}
\usepackage{amsmath}   

\usepackage{makecell}

\usepackage{xcolor}

\usepackage{amsthm} 

\usepackage{supertabular} 

\usepackage{multirow}
\newlength\savewidth

\begin{document}

\title{Marine Wireless Big Data: Efficient Transmission, Related Applications, and Challenges}
\author{
         Yuzhou~Li,
         Yu~Zhang,
         Wei~Li,
         and~Tao~Jiang
\thanks{The authors are with the School of Electronic Information and Communications, Huazhong University of Science and Technology, Wuhan, P.~R.~China.}
}
\maketitle
\IEEEpeerreviewmaketitle
\begin{abstract}
The vast volume of marine wireless sampling data and its continuously explosive growth herald the coming of the era of marine wireless big data. Two challenges imposed by these data are how to fast, reliably, and sustainably deliver them in extremely hostile marine environments and how to apply them after collection. In this article, we first propose an architecture of heterogeneous marine networks that flexibly exploits the existing underwater wireless techniques as a potential solution for fast data transmission. We then investigate the possibilities of and develop the schemes for energy-efficient and reliable undersea transmission without or slightly with data rate reduction. After discussing the data transmission, we summarize the possible applications of the collected big data and particularly focus on the problems of applying these data in sea-surface object detection and marine object recognition. Open issues and challenges that need to be further explored regarding transmission and detection/recognition are also discussed in the article.
\end{abstract}

\section{Introduction} \label{Section I}
The sea, as the cradle of life on earth, is an important space for human's survival and sustainable development. In recent years, the human beings have paid ever-increasing attentions on the marine exploration to establish a digital ocean. From a latest report, several marine observation projects, e.g., Argo, GOOS, OOI, IOOS, and NEPTUNE-Canada, have been launched for attempting to demystify the mysterious undersea world \cite{MarineBigData_AdvancesApplications_MPE2015}. Resorting to the 3S technologies, i.e., remote Sensing, geography information System, and global positioning System, the collected marine wireless data has amounted to the magnitude of PB per year, and is keeping increasing exponentially. For example, according to the statistics from the National Marine Information Center, China's annual marine data volume from remote sensing and collected marine data items in 2014 exceeded 41000 GB and 322 million, respectively \cite{Website_COIN}.
Another example comes from the National Oceanic and Atmospheric Administration in the United States, whose data amount per year had been up to 30 PB by the end of 2012 and over 3.5 million observational files were gathered together each day \cite{MarineBigData_AdvancesApplications_MPE2015,Website_NOAA}.
The vast volume of marine wireless data and its irresistible growth have announced the era of marine wireless big data has come.

Generally, the marine big data possesses ``4V'' features, defined as volume, variety, value, and velocity, \cite{bigdatafeatures_ieeeWC2017,bigdatafeatures_ieeeComputer2016}.
Specifically, due to myriad sources, sampling methods, and applications, the obtained marine wireless data is usually saved in different types such as audios, images, and videos, and thus is of high variety. Moreover, the original marine data usually has a low value relative to its volume (e.g., due to hydrology noises) and thus is low value density, which makes it hard to extract valuable information from it. Furthermore, the marine big data sometimes is needed to be processed at a high velocity to meet the requirements of some applications, such as the weather forecast. These ``4V'' characteristics impose tremendous challenges on designing efficient data-transmitting and data-analyzing techniques. In this article, we particularly focus on the volume and value features. More specifically, two important questions that arise for these large amounts of marine wireless sampling data are how to fast, reliably, and sustainably deliver them to collection centers in complex marine environments and what they can be used for.

Regarding transmission, acoustic communications have emerged as the most prominent choice in underwater scenarios so far. However, fast, energy-efficient, and reliable data transmission over the undersea acoustic medium is not easy due to the following facts.
\begin{itemize}
  \item\textbf{Violent path attenuations}. Underwater sound waves suffer from absorption loss (which increases violently with distance and frequency) and severe ambient noise.

  \item\textbf{Severe multiple path}. The reflections from the ocean surface and floor together with the varying reflections and refractions from virtual layers due to their variable conductivity produce severe multi-path propagation.
  \item\textbf{Large propagation delay}. Sound propagates underwater at a very low speed of about 1500~m/s, leading to a delay spreading over tens or even hundreds of milliseconds.
  \item\textbf{Strong Doppler effects}. The relative motion of transmitters and receivers resulted from the ocean dynamics creates extreme Doppler effects.
\end{itemize}
Hence, the undersea acoustic channel combines the worst aspects of poor quality of terrestrial mobile and high latency of satellite radio channels. As a result, it is rather energy- and spectrum-expensive to continuously satisfy the high-data-rate and reliable transmission demands. Even worse, most of the undersea devices have limited battery storage without effective energy replenishment means. As a consequence, it is well-worth investigating how to achieve energy-efficient and reliable transmission without or slightly sacrificing the data rate when delivering the marine big data.

Once the marine wireless big data has been collected after transmission, then the problem becomes how to use them for practical applications, e.g., supporting civil or military applications like resource exploration and marine rescue. Among them, two important are marine object detection and recognition. Specifically, marine object detection devotes to discovering objects far away leveraging devices like radar, which can find wide applications. For example, judging sea-skimming aircrafts in sea wars, avoiding icebergs and turbulence when sailing, finding rescuing targets in marine accidents, and so on.
Marine object recognition is to identify specific targets from collected information (such as identifying objects in an image or video sequence) by analyzing their unique features.
With the help of the collected marine big data, learning-based intelligent detection and recognition algorithms can be developed. In particular, this article focuses on the applications of marine big data in detecting sea-surface small targets and recognizing marine objects from images.

The remainder of this article is organized as follows. In Section~\ref{Section II}, we first propose an architecture of heterogeneous marine networks as a potential solution for fast wireless data transmission and then investigate the possibilities of fast, reliably, and sustainably delivering the marine big wireless sampling data. Section~\ref{Section III} first introduces potential applications of the collected marine big data, and then focuses on how to apply them in marine object detection and recognition. Finally, we conclude the article in Section~\ref {Section IV}.

\section{Marine Wireless Big Data Transmission} \label{Section II}

For the large volume of wireless sampling data, the first critical problem is how to reliably and fast deliver them to data centers in a sustainable manner. To achieve these aims, we first summarize the existing undersea wireless technologies and propose an architecture of heterogeneous undersea networks to make full use of them. We then investigate the possibility of energy-efficient undersea transmission without transmit rate reduction, as undersea devices usually cannot be conveniently energy-supplied. We also investigate the possibility of reliable undersea transmission that uses as few spectrums as possible for channel estimation and equalization (and thus more spectrums can be used for data transmission). Finally, we discuss the open issues and challenges when designing energy- and spectrum-efficient transmission schemes.

\begin{table*}[t]
      	\centering
      	\caption{Comparison among undersea wireless communication technologies}
      	\begin{tabular}{|c|c|c|c|c|}
      		\hline
      		\bfseries \makecell[tc]{Medium types} & \bfseries \makecell[tc]{Propagation speed} & \bfseries \makecell[tc]{Data rate} & \bfseries \makecell[tc]{Communication range} & \bfseries \makecell[tc]{Application scenarios}\\
      		\hline
      		\makecell[c]{EM} & $3.33\times10^7$ m/s & $\thicksim$Mb & $\leq10$ m & \makecell[c]{shallow water; localized network;\\cross air/water interface.}\\
      		\hline
           Acoustic & 1500 m/s & $\thicksim$Kb & $\thicksim$20 km & \makecell[c]{long range; small volume data transmission.}\\
      		\hline
           Optical & $3.33\times10^7$ m/s & $\thicksim$Gb & 10-100 m & \makecell[c]{clear water; line-of-sight;\\ real-time transmission.}\\
      		\hline
           MI & $3.33\times10^7$ m/s & $\thicksim$Mb & 10-100 m & \makecell[c]{oil reservoirs; water pipelines.}\\
      		\hline
      	\end{tabular}
      	\label{Table:ComparisionAmongWirelessTechnologies}
\end{table*}

\subsection{Heterogeneous Marine Networks} \label{Section II-A}
Undersea transmission and networks mainly adopt wireless technologies due to the difficulties in laying wired mediums such as cables and fiber. According to the adopted mediums, the current undersea wireless technologies can be classified to be electromagnetic (EM), acoustic, optical, and magnetic-induced (MI). Table~\ref{Table:ComparisionAmongWirelessTechnologies} summarizes the key features of these four technologies, where it is observed that each wireless technology has its own pros and cons.

\begin{figure}[t]
\centering \leavevmode \epsfxsize=3.5in  \epsfbox{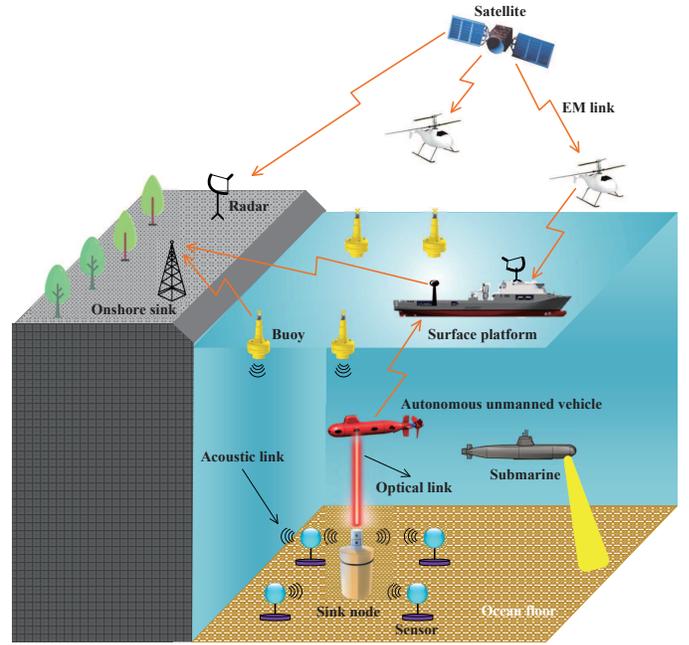}
\centering \caption{Architecture of heterogeneous marine networks.}
\label{Fig:HeterogeneousMarineNetworks}
\end{figure}

To boost the data rate, the network therefore should avoid their limitations and exploit their advantages as fully as possible. In this view, we propose an architecture of heterogeneous marine networks, which integrates and attempts to flexibly exploit all of them, as a potential solution for fast data transmission, as shown in Fig.~\ref{Fig:HeterogeneousMarineNetworks}. Its main components are described as follows.

\begin{itemize}
  \item \textbf{Above the sea surface}. In areas above the sea surface, the EM wave suffers negligible absorption loss, and achieves high data rate (in the magnitude of Mb) and low propagation delay and thus is the best choice.
  \item \textbf{Across the air/sea bound}. The EM wave is also a suitable choice in these areas, because it can smoothly pass the sea surface and extend the transmission range by following a path with least resistance when crossing the air/sea bound.
  \item \textbf{Under the sea}. In undersea environments, the acoustic can cover long distances with relatively low data rate and high delay, while the EM/optical/MI with high date rate is only suitable for short-range transmission. Therefore, no single technology can simultaneously satisfies the high-data-rate, long-range, and low-delay requirements. Our idea is that the undersea wireless devices first should be multi-mode-enabled if possible (might at the cost of other indexes such as the size) and thus they are able to simultaneously access multiple marine networks, e.g., optical and acoustic networks. The architecture then should be capable of flexibly exploiting heterogeneous mediums/networks to send and receive data concurrently through multiple of them or through the best one according to the application requirements. By this, network loads can be well balanced and cross-network resource optimization can also be introduced to further improve the transmission data rate.
\end{itemize}

Due to the predominance of underwater acoustic technologies, the remaining of the article mainly focuses on the problems whether it is possible to achieve sustainable, fast, and reliable underwater acoustic transmission under this heterogeneous architecture.

\subsection{Relay-Aided Energy-Efficient Transmission Schemes} \label{Section II-C}

The majority of undersea wireless devices are battery-limited without convenient supply, which necessitates prolonging their (and thus networks') lifetime to support continuously high-speed data transmission. From the evaluation results in \cite{UnderwaterAcousticChannel_CapacityOnDistance_acmSMCCR2007}, to achieve a given target signal-to-noise-ratio (SNR), the effective bandwidth decreases and the required transmit power increases both exponentially when the transmission distance grows. For example, when communicating with a device 1~km away at the SNR of 20~dB, the required bandwidth and transmit power are dozens of kHz and less than 1~W, respectively. However, the bandwidth becomes only about 1~kHz and the transmit power jumps to about 30~W when the distance is 100~km. These facts, i.e., bandwidth-range and power-range dependent features of the underwater acoustic channel, imply that shortening the hop distance by deploying relays along an underwater acoustic link may improve the transmission performance. Therefore, our idea for sustainably delivering the big sampling data as fast as possible is to devise relay-aided energy-efficient transmission schemes, which cannot only substantially reduce energy consumption but also maintain or even increase the data rate (because the effective bandwidth is expanded due to the decreased hop distance).

\begin{figure}[t]
  \centering
  \subfigure[End-to-end delay vs. number of relays]{
    \label{Fig:Delay_VS_RelayNodes} 
    \centering \leavevmode \epsfxsize=3.5in  \epsfbox{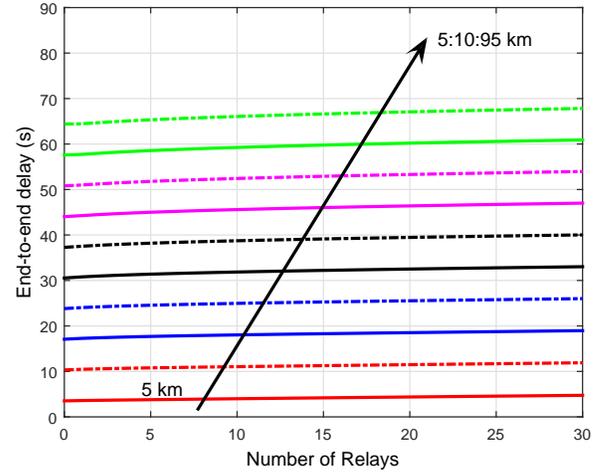}}
  \hspace{0in}
  \subfigure[Overall energy consumption vs. number of relays]{
    \label{Fig:Energy_VS_RelayNodes} 
    \centering \leavevmode \epsfxsize=3.5in  \epsfbox{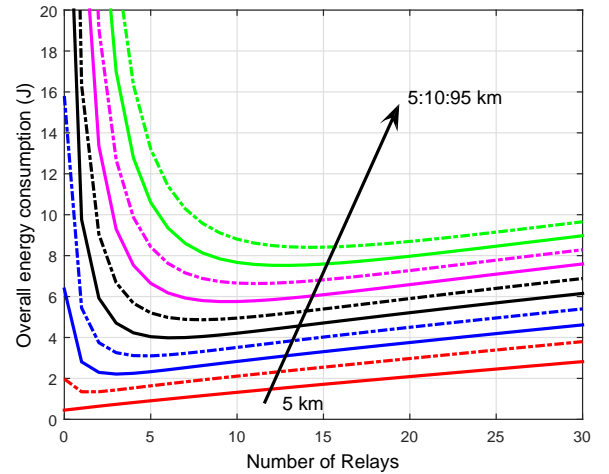}}
  \caption{Delay and consumed energy under different transmission distances between the transmitter and the receiver. In this figure, we use the same path loss and ambient noise models as in \cite{UnderwaterAcousticChannel_CapacityOnDistance_acmSMCCR2007}, where practical sound spreading, moderate shipping activity, and no wind situation are adopted. In addition, the modulation means, the target SNR, and the receive power are set to be BPSK, 10~dB, and 2.0~W, respectively.}
  \label{Fig:DelayEnergy_VS_RelayNodes}
\end{figure}

Despite the aforementioned advantages, relay deployment may in turn degrade the network performance as the deployed relays will spend extra time on packet forwarding and consume additional energy to receive data.
To address how relay deployment affects the performance of underwater acoustic networks, we consider linear underwater networks with all relays uniformly deployed on the line between the transmitter and receiver.
Fig.~\ref{Fig:Delay_VS_RelayNodes} and Fig.~\ref{Fig:Energy_VS_RelayNodes} exhibit how the end-to-end delay and the overall energy consumption vary with the number of relays, respectively.
From Fig.~\ref{Fig:Delay_VS_RelayNodes}, end-to-end delay curves are nearly constant for all transmission distances.
This is because the end-to-end delay is dominated by the propagation delay due to the low sound speed (about 1500 m/s underwater).
From Fig.~\ref{Fig:Energy_VS_RelayNodes}, when the distance between the transmitter and the receiver exceeding a threshold, deploying an appropriate number of relays indeed reduces the energy consumption, especially for those long-range transmission.
Furthermore, our obtained results show that compared with direct transmission, deploying a relay at the middle point can dramatically reduce the network energy consumption (up to 71.77\%) almost without increasing the end-to-end delay (less than 1.54\%) under typical parameter settings.
Integrating these facts, we can conclude that relay-aided transmission schemes cannot only significantly reduce the network energy consumption but also almost maintain a stable end-to-end delay in undersea acoustic networks.

\begin{figure}[t]
\centering \leavevmode \epsfxsize=3.5in  \epsfbox{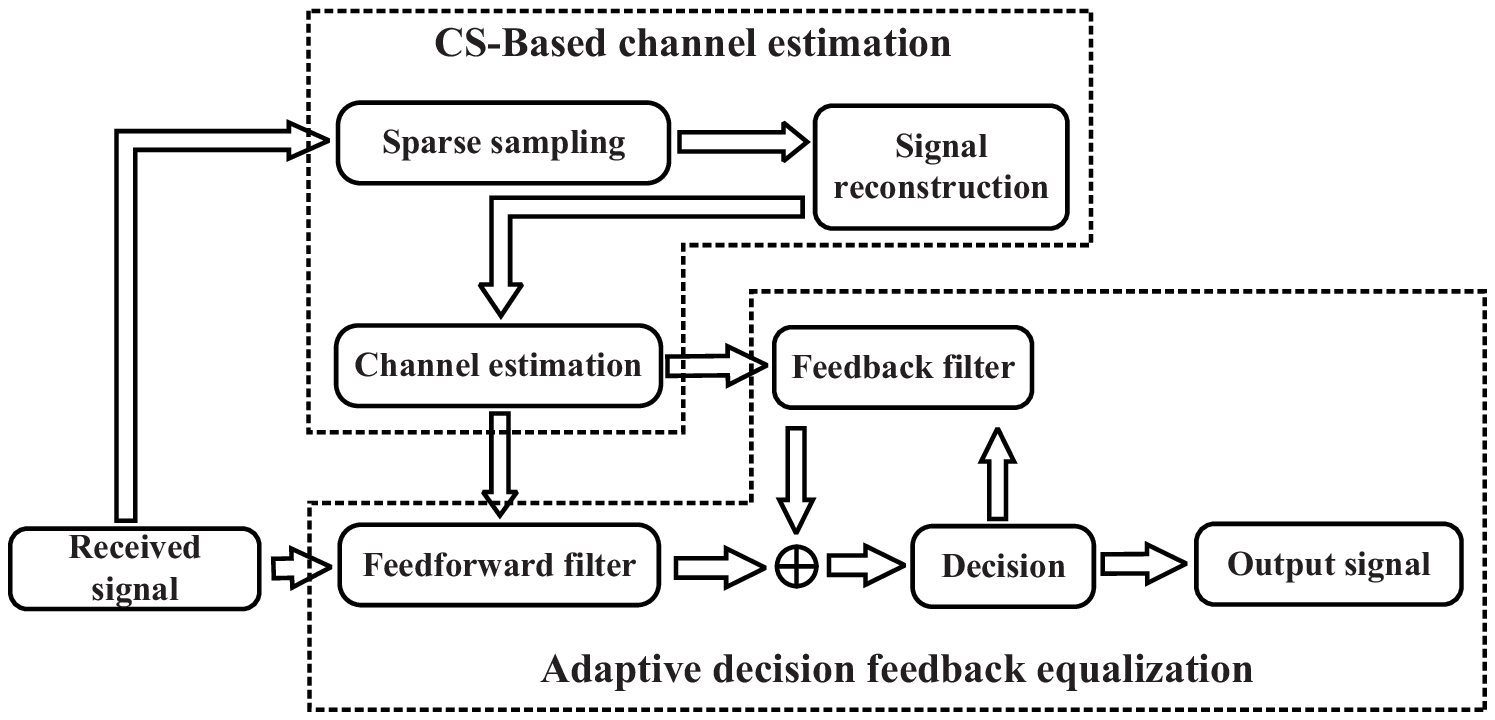}
\centering \caption{Adaptive decision feedback equalization combined with channel estimation.}
\label{Fig:ChannelEstimationAndADFE}
\end{figure}

\subsection{Spectrum-Efficient Reliable Transmission Schemes} \label{Section II-D}

Apart from energy efficiency, it is also important to guarantee the reliability of data transmission using as few pilots as possible (and thus more spectrums can be used to improve the data rate). It is known that accurately estimating acoustic channel parameters is crucial for reliable transmission. However, using few pilots to do this is not easy as underwater acoustic channels are significantly affected by underwater obstacles, water mobility, depth, temperature, salinity, acidity, and density, which usually vary in both spatial and temporal domains \cite{UnderwaterAcousticChannel_ChannelModel_ieeeCM2009}. There has been extensive investigation on channel estimation in terrestrial wireless systems by methods such as least squares (LS), minimum mean square error (MMSE), and maximum likelihood (ML). Nevertheless, these methods are originally designed for terrestrial dense multipath channels and thus reference signals are required, which inevitably wastes the very limited bandwidth of acoustic channels since reference signals carry no data information.

As a consequence, it is of significance to develop specialized techniques for underwater acoustic channel estimation that do not need to consume excessive spectrums for reference signals. By analyzing the collected big channel sampling data, it is found that, different from terrestrial dense multipath channels, underwater acoustic channels are sparse, with less than 10\% multipath channels occupying more than 85\% of the energy \cite{CompressiveSampling_PICM2006}. This sparsity indicates that, instead of first sampling the underwater channel state information at a high rate and then compressing the sampled data, we can directly sense the data in a compressed form with little information loss, i.e., compression sensing (CS) at a lower sampling rate.
With less sampling data, CS can decrease the number of training sequences or pilots, and thus leave more bandwidth for underwater data delivering.
Moreover, CS typically adopts optimization methods, such as tractable mixed-norm optimization programs, efficient greedy algorithms, or fast iterative thresholding methods, to reliably reconstruct the channel state signals from the compressed data.
In a nutshell, we can exploit the CS theory to devise spectrum-friendly underwater acoustic channel estimation methods.

However, reliability cannot be guaranteed when channel estimation does not work well, which possibly often happens in the extremely hostile underwater environments. It is thus also needed to mitigate the inter symbol interference (ISI) caused by the large multipath delay spread at receivers. Attributed to the ability in automatically tracking the time-varying channel parameters, the adaptive decision feedback equalizer (ADFE) is a good choice for the ISI mitigation for underwater acoustic channels \cite{WirelessCom_Smith}. However, the initialization of the ADFE requires lots of training sequences that are embedded in data packets, which unavoidably occupy a large number of scarce acoustic spectrums. Considering that channel estimation can perceive frequency responses of channels, our proposed scheme for spectrum-efficient reliable data transmission is by combining the spectrum-friendly CS and the ISI-resistant ADFE, as shown in Fig.~\ref{Fig:ChannelEstimationAndADFE}. In this scheme, the CS-based channel estimation module first samples and then reconstructs underwater channel responses. With these priori-knowledge, the feedforward filter first tracks the variation of the acoustic channel with less training sequences. After that, the other parts of the ADFE adaptively eliminate the channel distortion and the ISI. By this scheme, data rate can be improved (because more spectrums are reserved for data delivering) without degrading reliability.

\subsection{Open Issues and Challenges} \label{Section II-E}
Currently, investigation on relay deployment, channel estimation, and equalizer design usually adopts ideal assumptions and simplified models, which makes the proposed schemes can hardly be applied to realistic underwater networks. Some of problems that remain open are summarized as follows.
\begin{itemize}
  \item Existing works explore the relay deployment problems mainly assuming that sound propagates straightly underwater, but this is actually not right. In practice, the acoustic propagation path is bent due to the different temperature, pressure, salinity, and density along vertical layers. Then, it is very challenging to determine whether to deploy relays and where to deploy them for reducing energy cost without sacrificing data rate. This becomes much more complicated when we consider realistic underwater acoustic channel models, which are very complex and thus simplified in existing works.
  \item Unlike terrestrial radio propagation models, acoustic channel parameters are different in vertical layers. This makes channel estimation very challenging when an acoustic path goes across multiple vertical layers. On the other hand, the ADFE usually needs a large number of taps to guarantee the performance of the ISI mitigation when a high data rate is also required, resulting in high computational complexity. Moreover, it is also difficult to accurately and quickly determine tap coefficients of the ADFE in rapidly time-varying acoustic channels.
\end{itemize}

\section{Marine Wireless Big Data for Applications} \label{Section III}
In this section, we first introduce the overall possible applications of the collected marine wireless big data and then particularly focus on its applications in marine object detection and recognition.

\subsection{Overall Applications of Marine Wireless Big Data} \label{Section III-A}
The values of the collected marine wireless big data lie in that it can be directly utilized or deeply mined to explore, develop, and protect the mysterious marine world. Some of its important applications are listed as follows.

\begin{itemize}
  \item \textbf{Marine exploration}. Observation data records abundant information about ocean geosciences, physics, chemistry, and biology, and also provides databases for sea mapping, current modeling, and species discovery. For example, analyzing the Argo data finds that the earth is seeking an intensification of global hydrological cycle \cite{MarinePhenomena_GlobalWaterCycle_Science2012}.
  \item{\textbf{Marine forecast and warning system}}. Maritime activities are threatened by frequently-changed climate and potential disasters such as typhoon, tsunami, and undersea earthquake. Marine big data can be mined to provide timely and wide-ranged weather forecast and disaster warning information. For instance, the Neptune project devotes to forecasting undersea earthquake and tsunami by analyzing observation data including seismic activity, faulting activity, and mid-oceanic ridges \cite{MarineProject_NEPTUNE_aguFMA2009,MarinePhenomena_ChileanTsunami_PAG2013}.
  \item{\textbf{Marine object detection and recognition}}. Marine object detection and recognition have wide applications in warship tracking, autonomous navigation, marine rescue, and etc. Data-driven machine learning techniques enable us to design efficient detectors and classifiers.
  \item{\textbf{Marine monitoring}}. Lots of large-scale infrastructures and platforms have been built on/beneath the sea surface to explore the ocean. Their real-time operational states can be monitored without being there by analyzing the big data sampled by numerous embedded sensors.
  \item{\textbf{Marine protection}}. Destructive damages to marine environments and ecology will be incurred by anti-natural development, which can be avoided under the guidance of ocean rules mined from the marine big data.
\end{itemize}

In what follows, we discuss how to apply the collected marine wireless big data to detect and recognize marine objects in detail.

\begin{figure}[t]
\centering \leavevmode \epsfxsize=3.5in  \epsfbox{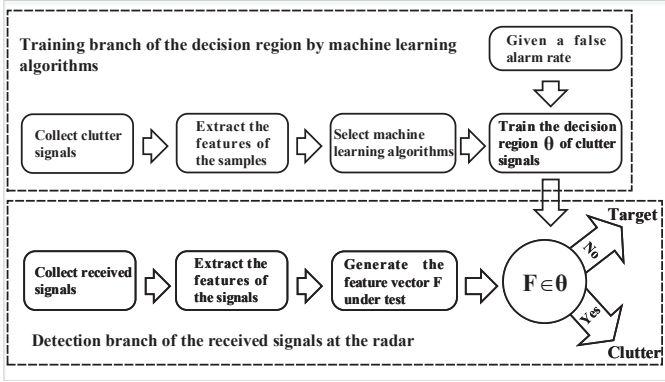}
\centering \caption{Procedure of data-driven machine learning algorithms to detect marine targets from sea clutter.} \label{Fig:MarineObjectDetection}
\end{figure}

\subsection{Marine Object Detection} \label{Section III-B}
Marine big data can be used to assist the marine surface surveillance radar to find sea-surface moving small targets, such as small boats, buoys, or even a person in the sea. Specifically, these data is analyzed by the radar to determine a proper decision region for an accurate target detection. However, it is difficult to extract the weak target signals from the data of chaotic received signals due to the effects of the strong spiky components of sea clutter at a low grazing angle. Considering the impacts of the sea clutter, adaptive methods for small target detection based on statistical sea clutter models have been widely adopted \cite{FloatingSmallTargetsDetection_TriFeature_ieeeTAES2014}. These models are usually established as compound-Gaussian models \cite{SeaClutter_NonGaussianMitigation_ieeeJOE2004,SeaClutter_GaussianPerformanceAnalysis_ieeeTAES2002}, obtained by analyzing the big existing datasets, such as the IPIX radar data, where each dataset includes 14 consecutive range cells with the length of time series at each cell being $2^{17}$ and contains four polarization modes (i.e., HH, HV, VH, and VV). When applying adaptive methods such as adaptive threshold techniques, these obtained models can be seen as the prior-knowledge to dynamically estimate the background level of the sea clutter and thus can be used to update the decision threshold or the decision region.

Nevertheless, conventional adaptive detectors, such as the likelihood ratio test, matched filter, and Bayesian detection, can only be applied under the condition that the sea clutter and target signals are separate enough in the Doppler domain. However, this precondition usually does not hold for marine objects because of their Doppler effects incurred by the ocean dynamics, especially in cases of detecting maneuvering targets. A promising solution is to separate the sea clutter and target signals by their inherent features, but how to define and measure these features is still unknown. This might become possible when we have collected a large volume of detection data that includes sea clutter and target signals, since we may find some key features for signal separation by analyzing them. For example, after analyzing the IPIX radar data, it is found that
there are significant differences between the received signals at clutter-only cell and target cell in amplitude.
Based on this fact, a feature referred to as relative average amplitude (RAA), defined as the ratio of the average amplitudes at the cell under test (CUT) to that of the reference cells around the CUT, has been extracted to detect target signals from sea clutter signals~\cite{FloatingSmallTargetsDetection_TriFeature_ieeeTAES2014}.

One important method for feature extraction from the collected big detection data and intelligent decision region determination is machine learning.
The feasibility of the decision region determination depends on two factors: the various types of sea states and the computational complexity of learning algorithms. Thus, the challenging problem is converted to separate the target signals from the sea clutter signals with low computational complexity at the certain type of the sea states. Particularly, the procedure of data-driven machine learning algorithms to detect marine targets can be divided into three steps, as shown in Fig.~\ref{Fig:MarineObjectDetection}. Firstly, collect clutter signal data from the marine surface surveillance radars and mine the features from it. Then, train the decision region of the clutter pattern by machine learning algorithms at a given false alarm rate.
For instance, use support vector machine (SVM) to train a non-probabilistic binary linear classifier, or apply decision tree (DT) to select the top influential features and to construct a tree-like decision graph.
Finally, use the trained decision region $\theta$ to test the received signals and decide whether there is a target or not.

\begin{figure*}[t]
    	\centering
    	\includegraphics{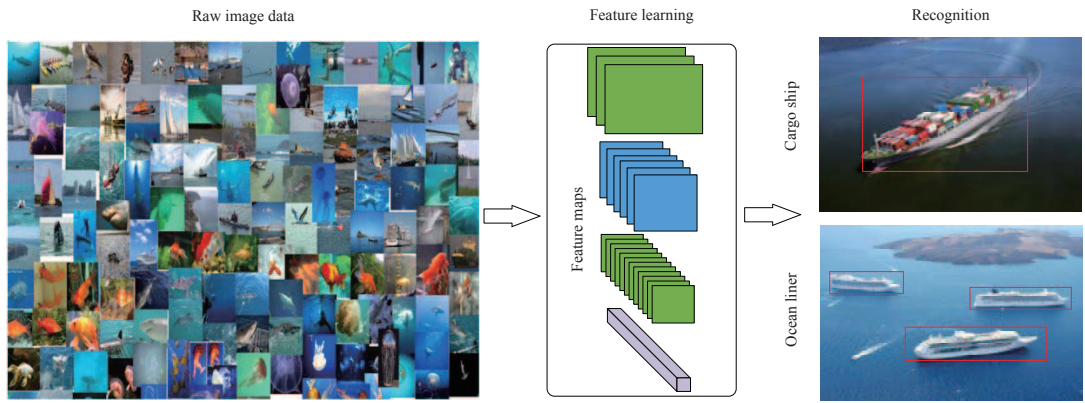}
    	\caption{An example of a data-driven deep learning algorithm for marine object recognition.}
    	\label{Fig:MarineObjectRecognition}
\end{figure*}

\subsection{Marine Object Recognition} \label{Section III-C}
Another important application is to identify specific objects from collected videos,  images, and other types of information sampled by cameras, radars, and other sensors, i.e., marine object recognition, by extracting and comparing their features. Different from colorful and variable terrestrial environments, the marine background is relatively monotonous in both color and contents because the objects are generally surrounded by vast ambient water or empty sky. This feature facilitates to separate the objects from the background and to extract the features of objects. On the other hand, due to the dramatically changed hydrological conditions and marine atmosphere, the underwater channel qualities are both temporally and spatially varying. As a result, the collected object information is usually incomplete, noisy, and multi-modal, which makes it difficult to accurately recognize the objects. Traditional marine object recognition methods (such as chaos and fractal theories, wavelet analysis) require manually designed features or a large number of priori parameters to learn features. However, these manually designed features or parameters can hardly be used to distinguish enough kinds of objects and are usually not immune to the hostile marine environments.
It is thus important to develop specialized methods to obtain a series of distinguished and environment-immune features for recognizing different objects.

As the collected data contains the information of both the objects and the marine environments, data-driven feature learning methods have been proposed to let the recognition system automatically choose the distinctive features that can be mined from the marine big data. Taking deep learning as an example, it typically consists of multiple layers for feature extracting. The input layer directly learn features from the original data, and the deeper layers can extract more abstract high-level features of the objects by combining the low-layer features. Through backpropagation or other optimization algorithms, deep learning can adaptively adjust the training parameters within each layer (i.e., choose the best combination of different layers' features) to achieve the optimal recognition accuracy. Fig. 5 shows an example of using a deep learning algorithm to recognize marine objects in images, where the algorithm first learns from the large amount of image data to extract key features and then identifies the targets by labeling them as the name with the highest recognition probability. The obtained results show an excellent recognition performance.

\subsection{Open Issues and Challenges} \label{Section III-D}
Although extensive research has been conducted on the marine object detection and recognition, it is still far away from accuracy, rapidness, and cost-efficiency. Some of key challenging problems are summarized as follows.
\begin{itemize}
\item Extracting key features from the collected wireless detection data is crucial for accurate marine object detection. However, we usually have no ideas on how to define such features with physical implications. What's worse, single feature is often insufficient for accurately detecting marine objects, especially detecting those floating or maneuvering small targets. Thus, how to find a multiple-dimensional feature space to distinguish clutter and object signals is another challenge.
\item It remains open to effectively extract distinctive parameters and precisely evaluate them for a set of objects from the marine image data, since multiple similarities may exist in different marine objects, e.g., various on-board equipments (such as masts, radars, and transceivers) usually have similar rim shapes. In addition, to improve the accuracy of object recognition, it is a trend to deepen the layers of artificial neural networks, which in turn imposes computational burdens on the system. Hence, how to strike a flexible balance between accuracy and complexity also needs to be further investigated.
\end{itemize}	

\section{Conclusions} \label{Section IV}
Establishing a digital ocean for comprehensive ocean exploitation has drawn more and more attentions across the world, which promotes the coming of the era of marine wireless big data. The ``4V'' characteristics of the marine wireless big data have imposed challenges on the data transmission and analysis. Among them, this article has investigated two important problems that are how to timely and reliably transmit the large volume of marine sampling data in complex ocean environments and how to apply the collected marine big data into practical applications. Specifically, we have first integrated the existing marine wireless technologies to propose an architecture of heterogeneous marine networks for timely delivering marine wireless big data. Due to the requirements on energy conservation and transmission reliability, we have then discussed whether it is possible to energy-efficiently and reliably to deliver marine big data without the cost of data rate. Furthermore, we have summarized the potential applications of these collected marine data and particularly concentrated on two of them, namely the marine object detection and recognition. In addition, this article has also covered the problems and challenges that remain open from the aspects of transmission and applications.

\section{Acknowledgement}
This work was supported in part by the National Science Foundation of China with Grant numbers 61601192, 61601193, 61729101, and 61631015, the Major Program of National Natural Science Foundation of Hubei in China with Grant 2016CFA009, and the Fundamental Research Funds for the Central Universities with Grant 2016YXMS298.

\bibliographystyle{IEEEtran}
\bibliography{IEEEabrv,MarineBigData_Reference}

\section*{Biographies}

{\footnotesize{\noindent Yuzhou~Li [M'14] (yuzhouli@hust.edu.cn) received the Ph.D. degree in communications and information systems from the School of Telecommunications Engineering, Xidian University, Xi'an, China, in December 2015. Since then, he has been with the School of Electronic Information and Communications, Huazhong University of Science and Technology, Wuhan, China, where he is currently an Assistant Professor. His research interests include 5G wireless networks, marine object detection and recognition, and undersea localization.}}

\vspace{1em}

{\footnotesize{\noindent Yu~Zhang (yu$\_$zhang@hust.edu.cn) received the B.Eng. degree in communications engineering from the School
of Electronic Information and Communications, Huazhong University of Science and Technology, Wuhan, China, where he is currently pursuing the M.S. degree. His research interests include marine information networks and underwater communications.}}

\vspace{1em}

{\footnotesize{\noindent Wei~Li (autolee@hust.edu.cn) received the B.Eng. degree in control engineering from the School of Automation, Hubei University of Science and Technology, Hubei, China. Now he is currently pursuing the Ph.D degree in Wuhan National Laboratory for Optoelectronics. His research interests include underwater sensor networks and underwater localization.}}

\vspace{1em}

{\footnotesize{\noindent Tao~Jiang [M'06-SM'10] (taojiang@hust.edu.cn) is currently a Distinguished Professor with the School of Electronics Information and Communications, Huazhong University of Science and Technology, Wuhan, China. He has authored or co-authored over 300 technical papers and 5 books in the areas of wireless communications and networks. He is the associate editor-in-chief of China Communications and on the Editorial Board of IEEE Transactions on Signal Processing and on Vehicular Technology, among others.}}

\vspace{1em}

\end{document}